# Recrystallization of epitaxial GaN under indentation

S. Dhara,[1,a)] C. R. Das,[2] H. C. Hsu,[3] Baldev Raj,[2] A. K. Bhaduri,[2] L. C. Chen,[3] K. H. Chen,[4] S. K. Albert,[2] and Ayan Ray[5]

[1] Department of Electrical Engineering, Institute for Innovations and Advanced Studies, National Cheng Kung University, Tainan-701, Taiwan.

[2] Metallurgy and Materials Group, Indira Gandhi Center for Atomic Research, Kalpakkam-603102, India

[3] Center for Condensed Matter Sciences, National Taiwan University, Taipei-106, Taiwan

[4] Institute of Atomic and Molecular Sciences, Academia Sinica, Taipei-106, Taiwan

[5] Department of Physics, Indian Institute of Technology, Kharagpur- 721 302

We report recrystallization of epitaxial (epi-) GaN(0001) film under indentation. Hardness value is measured ~10 GPa, using a Berkovich indenter. 'Pop-in' burst in the loading line indicates nucleation of dislocations setting in plastic motion of lattice atoms under stress field for the recrystallization process. Micro-Raman studies are used to identify the recrystallization process. Raman area mapping indicates the crystallized region. Phonon mode corresponding to $E_2$(high) ~570 cm$^{-1}$ in the as-grown epi-GaN is redshifted to stress free value ~ 567 cm$^{-1}$ in the indented region. Evolution of $A_1$(TO) and $E_1$(TO) phonon modes are also reported to signify the recrystallization process.

[*] Author to whom correspondence should be addressed; electronic mail: dhara@igcar.gov.in

[a)] Part of the work is performed in Institute of Atomic and Molecular Sciences, Academia Sinica, Taipei-106, Taiwan. Presently on leave from Materials Science Division, Indira Gandhi Center for Atomic Research, Kalpakkam-603102, India

Growth of epitaxial (epi-) GaN film with low dislocation density either by pre-grooving the buffer layer[1] or by lateral epitaxial overgrowth (LEO)[2] is basically to reduce residual strain in the system and help crystallization process. A remarkably low threading dislocation (TD) density ~$10^7$ cm$^{-2}$ is reported in epi-GaN film adopting this technique. GaN is one of the most important optoelectronic semiconductors for its applications as white light source, blue diode leading to UV lithography.[3] High dislocation density and related issues in GaN are one among few of the major hindrances for the application as blue laser.[4] Thus, removal of stress is one of the prime objectives for opto-electronic device applications in GaN. A combination of biaxial and hydrostatic stresses originating form dislocation related extended defects and point defects, respectively, are reported generally in epi-GaN.[5] However, hydrostatic stress alone is estimated inside the indented region of epi-GaN without implication of any structural changes.[6]

While the above said techniques[1,2] are well reported for the growth of epi-GaN with reduced dislocation density, we report here the recrystallization of epi-GaN under indentation with various loads and loading-unloading rates. Micro-Raman spectra along with area mapping, for the spectral region of interest, are studied using 514.5 nm excitation of Ar$^+$ laser.

The undoped wurtzite (0001) oriented GaN epi-layer of 6 μm thickness, grown by MOCVD on a (0001) oriented crystalline sapphire substrate with TD <$5\times10^8$ cm$^{-2}$ (TDI, USA), is used in the present study. The sample is indented using the micro-indenter with a Berkovich diamond indenter (three-sided pyramid with a nominal tip radius of 50 nm and having centerline-to-face angle, Ψ=65.3º). The indentation conditions are as follows: load 30-500 mN; loading-unloading rate 0.1-50 mN.s$^{-1}$, and holding time 5 s. It is worth to mention that owing to reliability of data, low loading-unloading rate ~ 01.-0.5 mN.s$^{-1}$ is not achievable for loads >100 mN and high loading-unloading ~ 50 mN.s$^{-1}$ is limited for loads ≥ 350 mN. Loads are chosen such that penetration depth is ≤ 20% of the sample thickness, so as to avoid ambiguity for the substrate contribution. The optical image of the indentation is taken using microscope with 100X/0.9 objective (Olympus, BXFM). The length of the side of the triangle indentation is about 10 μm. A LabRam HR800 (Jobin-Yvon, France) spectrometer with an automatized XY-table of acquisition is used to record the Raman spectra with excitation wavelength of 632.8nm nm of He-Ne laser. Raman area mapping is also recorded for the spectral region of interest. Considering a laser spot size of 1 μm and the thickness of the GaN layer the probed volume is less than 1 μm$^3$. All the Raman spectra are recorded in backscattering geometry.

We extensively study the hardness [Fig.1(a)] and the reduced modulus [Fig.1(b)] of epi-GaN(0001) at various loads with different loading rates. Hardness value equivalent to bulk GaN (~10 GPa)[7] is achieved at high load ≥ 100 mN [Fig. 1(a)]. At a load of 100 mN, the bulk value is achieved only at low loading rate (≤ 0.5 mN.s$^{-1}$). As a matter of fact hardness increases with increasing loading rates in the low load regime. This goes well with the conventional wisdom of materials that at low loads, where defect formation is mainly close to the surface, increasing number of extended defects (dislocations) with increasing loading rate enhances the strength (hardness) of the material. Formation of extended defects can be explained with geometrically necessary dislocation model,[8] which in turn explains the depth dependent hardness variation for the crystalline material. Irrespective of loading rate, bulk hardness value is achieved steadily at higher load >100 mN. Different values of hardness for GaN are reported for different indentation geometry in the literature.[9] We need to state that the reliability of the data presented is not always upto mark, as contribution from sapphire substrate (hardness ~ 22 GPa)[10] is not always taken care properly. Indentation depth of ~ 800 nm is chosen using Berkovich indenter for GaN film thickness of ~2 μm on sapphire substrate.[8] The measured depth is ~ 40% of film thickness and can be regarded as not acceptable, as substrate effect is imminent in the reported value. However, hardness value of both bulk and epi-GaN is unanimously accepted ~10 GPa for most of other standard indenter.[9,11] Reduced modulus ~100 GPa is reported at high loads [Fig. 1(b)]. Typical optical microscopic image of the crack free triangular indent region without significant pile up is shown in the inset of Fig.1(b). The observed value of reduced modulus is realistically close to the reported Yong's modulus value of ~181 GPa for bulk GaN. On the other hand, a large range of Young's modulus is presented for epi-GaN depending on different loading conditions.[7,9,11] We must restate here that we have also observed high vales of elastic modulus > 200 GPa at low loads [Fig.1(b)], implying a proper choice of loading-unloading condition is important in reporting materials properties in this technique.

Typical loading-unloading curves are shown for low load [Fig. 2(a)] and high load [Fig. 2(b)] for a fixed loading-unloading rate of 10 mN.s$^{-1}$. In both cases, 'pop-in' burst in the loading line [encircled in Figs. 2(a) and 2(b)] as well as at the apex is noticed. The mechanism responsible for the 'pop-in' burst appears to be associated with the nucleation and movement of dislocation sources including lattice atoms for the possible crystallization process.[12]

The immediate implication in the structural transformation close to the indented region is studied using micro-Raman spectroscopy. $E_2$ (high) mode at 570 cm$^{-1}$, measured for a typical sample outside the indented region, resembles the reported value of epi-GaN on sapphire substrate [Fig. 3(a)]. Inset shows the spots measured out- and inside of the indentation mark recorded in the optical microscope attached to the spectrometer. However, micro-Raman measurements inside the indented region redshifts the peak gradually to 567 cm$^{-1}$ from close to the edge to the center of indented spot. This value is close to the calculated and measured value for $E_2$ (high) phonon of bulk GaN and GaN nanowire under stress free conditions.[13,14] Double peaks are observed for the region close to the edge of the indentation [Fig. 3(a)] showing contributions from both the stressed region outside and stress free region at the interface of the indented region. Assuming hydrostatic stress alone inside the indentation region,[6] a stress of <1 GPa is required for the 3 cm$^{-1}$ shift of $E_2$(high) peak position, as the reported pressure coefficient of the observed Raman peak is -3.55 cm$^{-1}$GPa$^{-1}$ for hydrostatic stress alone in GaN.[15] The distribution of pressure under the indenter given by[16]

$$p(r) = \frac{E}{2(1-v^2)} \frac{\cosh^{-1}(a/r)}{\tan \Psi}, 0 \leq r \leq a \quad \ldots\ldots\ldots\ldots\ldots\ldots\ldots\ldots\ldots (1)$$

where $E$ is Young's modulus, $v$ is Poisson's ratio (0.22 for epi-GaN),[11] $a$ is the contact radius, and $r$ is the radial coordinate in the surface. With $\Psi$ of 65.3° in the Berkovich indenter and using experimentally observed (in the most conservative approach) reduced modulus of 100 GPa, the pressure at the central region (for a spread of ~1 μm in the micro-Raman resolution at the centre) can be calculated (Eqn.1) as ~55 GPa (exactly at the centre) -10 GPa (at the boundary of ~1 μm spot at centre). Thus the estimated pressure requirement of <1 GPa for the observed Raman shift can be always provided in this technique. Raman imaging [Fig. 3(b)] using spectral part of 568-573 cm$^{-1}$ shows red (bright in the grayscale) region lying outside the indented region and 562-568 cm$^{-1}$ shows Green (bright in the grayscale) region lying inside the indented region. This observation proves, unambiguously, that the 567 cm$^{-1}$ signal originates from the stress free region inside the indented mark and 570 cm$^{-1}$ peak originates from the bulk of the sample. Thus release of stress, and thereby recrystallization of GaN, using indentation is further confirmed with the appearance of $A_1$(TO) and $E_1$(TO) modes at 531 cm$^{-1}$ and 559 cm$^{-1}$, respectively, in the indented region [Fig. 3(a)]. These values conform well to the reported values in the bulk GaN.[14] We may also state here that according to selection rule in wurtzite crystal of GaN, TO phonon modes are forbidden in the backscattering geometry for the (0001) oriented planes.[14] However, small misorientations of crystallites in the indented region allow phonon modes, corresponding to other crystalline orientations, to appear in the present scattering configuration. As a consequence, appearance of TO modes provides additional information of crystallinity in the indented region for other orientations containing different crystalline planes.

In conclusion, an indentation induced recrystallization is reported in epi-GaN using a Berkovich indenter. A typical hardness value of ~10 GPa is measured in the epi-GaN which conform well with reports using different other indenter geometrical configurations. Dislocation stress field might have been released at the centre of indentation region with the dislocation motion under the indentation stress. The amount of stress required for the recrystallization process is validated with the calculated value of stress in the indented volume.

We acknowledge H. C. Lo of CCMS, Taipei, Taiwan for his useful comments.

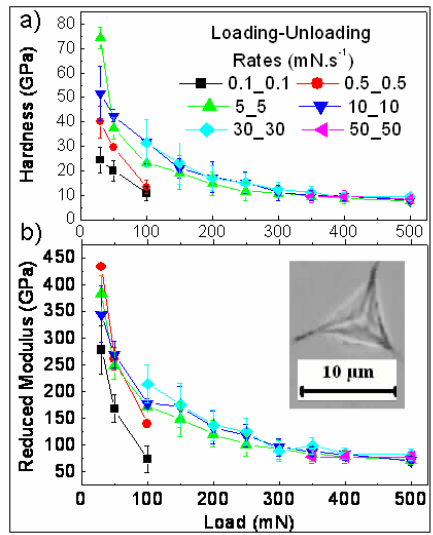

Fig. 1. Measured values of (a) hardness and (b) reduced modulus of epi-GaN(0001) with load. Loading-unloading rate is also varied for the detailed studies. Lines are joined as a guide to eyes. Inset in (b) shows the typical optical image of the crack free indent spot.

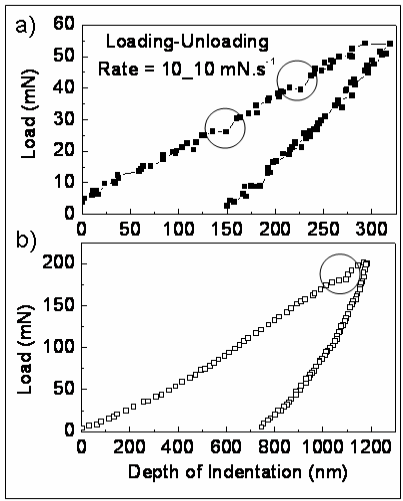

Fig. 2. Typical loading-unloading lines at a fixed loading-unloading rate of 10 mN.s$^{-1}$ for (a) low (50 mN) and (b) high (200 mN) loads. The encircled regions are called as 'pop-in' burst.

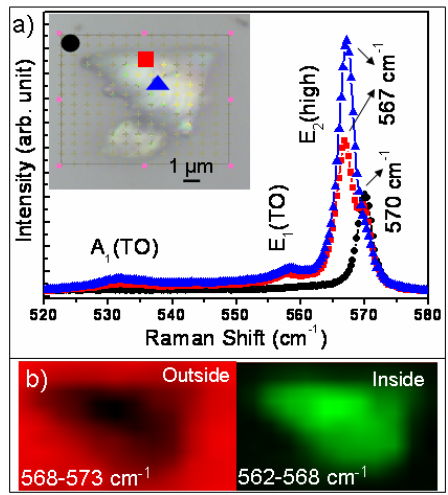

Fig. 3. (a) Micro-Raman spectra for epi-GaN outside and different regions inside the indentation spot. Inset shows corresponding optical image of the indentation spot. (b) Area mapping of outside and inside of the indentation spot using different spectral regions indicated in the picture.